\documentclass[pre,twocolumn,letterpaper,nofootinbib]{revtex4}

\usepackage[pdftex]{graphicx}
\usepackage{float}
\usepackage{tikz}
\usepackage{pgfplots}
\usepackage{pgfplotstable}
\usepackage{subfigure}
\usepackage{amsmath,tabularx}
\usepackage{amssymb}
\usepackage{array}
\usepackage{listings}
\usepackage{hyperref} 
\newcolumntype{C}[1]{>{\centering\arraybackslash}p{#1}}
\usetikzlibrary{arrows,automata}
\usetikzlibrary{arrows,shapes}
\newcommand{\nucrit}{\nu_\ast}
\newcommand{\nupred}{\nu}

\begin{document}

%\lstset{language=Matlab} 

\title{Symmetry-respecting real-space renormalization for the quantum Ashkin-Teller model}

\author{Aroon O'Brien}
\author{Stephen D.\ Bartlett}
\author{Andrew C.\ Doherty}
\author{Steven T.\ Flammia}
\affiliation{Centre for Engineered Quantum Systems, School of Physics, The University of Sydney, Sydney, NSW 2006, Australia}

\date{16 September 2015}

\begin{abstract}
We use a simple real-space renormalization group approach to investigate the critical behavior of the quantum Ashkin-Teller model, a one-dimensional quantum spin chain possessing a line of criticality along which critical exponents vary continuously. This approach, which is based on exploiting the on-site symmetry of the model, has been shown to be surprisingly accurate for predicting some aspects of the critical behavior of the quantum transverse-field Ising model.  Our investigation explores this approach in more generality, in a model where the critical behavior has a richer structure but which reduces to the simpler Ising case at a special point.  We demonstrate that the correlation length critical exponent as predicted from this real-space renormalization group approach is in broad agreement with the corresponding results from conformal field theory along the line of criticality.  Near the Ising special point, the error in the estimated critical exponent from this simple method is comparable to that of numerically-intensive simulations based on much more sophisticated methods, although the accuracy decreases away from the decoupled Ising model point. 
\end{abstract}

\maketitle

\section{Introduction}

Real-space renormalization group (RG) methods have a long and successful history in the study of critical behavior in quantum many-body systems~\cite{Goldenfeld,Efrati2014}.  Research into such methods remains very active, as new innovations seek a delicate balance:  to be tractable, the exponential size of the state space for large quantum many-body systems requires an extreme truncation of degrees of freedom, yet discarding too much can lead to inaccurate predictions.  Real-space RG methods beyond the original blocking approach of Kadanoff~\cite{Kadanoff} include the density matrix renormalization group~\cite{White1992}, projected entangled pair states~\cite{Verstraete2008}, the multiscale entanglement renormalization ansatz (MERA)~\cite{Vidal2007}, and tensor network renormalization~\cite{Evenbly2014}, among others.

In contrast to these sophisticated techniques, it has recently been shown that a very simple real-space RG map can be used to accurately predict some of the critical behavior of the quantum transverse-field Ising model~\cite{Miyazaki2011,Kubica}.  Rather than requiring intensive numerical calculations, this method can be solved analytically and gives closed-form expressions for RG fixed points and critical exponents.  This simple method exactly predicts the correlation-length critical exponent $\nu = 1$ for the one-dimensional model, and is surprisingly accurate for higher-dimensional models on a variety of lattices~\cite{Miyazaki2011,Kubica}.  The method is shown to yield accurate results for the Potts model as well~\cite{Kubica}.  It is unclear, however, whether this success is particular to the Ising model (and variants, like the Potts model) or can be applied to a wider range of models and more general studies of criticality.

In this paper, we formalize and generalize this simple real-space RG technique to preserve on-site symmetries under blocking; we refer to this method as the \emph{symmetry-respecting real-space renormalization group (SRS RG)}.  We then apply the SRS RG to the quantum Ashkin-Teller (AT) model~\cite{Solyom1981,Ashkin1943}.  This one-dimensional quantum spin lattice model has several properties that make it desirable for an investigation into the general efficacy of the SRS RG map.  First, it includes many features of the Ising model, such as the on-site symmetry that has been argued to be a key component of the success of this simple technique~\cite{Miyazaki2011,Kubica}.  In fact, for a particular choice of Hamiltonian parameters, the AT model reduces to two uncoupled Ising models.  The AT model has a much richer structure, however.  Most notably, it possesses a line of criticality, along which the correlation length critical exponent is expected to vary continuously.  Therefore, calculating this critical exponent using the SRS RG along this line is a useful test of the method's performance.  We note that the AT model along this line of criticality is equivalent to the exactly-solvable six-vertex model~\cite{Kohmoto}, and is expected to be described by a conformal field theory (CFT) with central charge $c=1$, specifically the so-called \emph{$c=1$ orbifold boson CFT}.  We can compare the calculated critical exponents from SRS RG with those obtained from the six-vertex model~\cite{Kohmoto} as well as the known behavior of the CFT.  A recent study of the critical behavior of the AT model and a detailed comparison with the $c=1$ orbifold boson CFT predictions was performed using MERA~\cite{Bridgeman2015} (a much more sophisticated and numerically-intensive real-space RG method), and this provides a useful benchmark.

Our analysis shows that this symmetry-respecting real-space RG scheme can quantitatively capture aspects of the critical behaviour for the AT model, and performs reasonably compared with more sophisticated numerical methods especially given its simplicity.  By focussing on preserving the symmetry at each blocking, the SRS RG captures many of the important features of the RG flow and behavior along the critical line.  Although this technique may be of limited use as an accurate numerical simulation method, its simplicity and analytical form may allow this method to play a role analogous to mean-field theory for critical RG flows.

Specifically, when applied to the AT model, we find that the SRS RG identifies fixed points on the critical line.  Although the scaling field associated with flow along this line is predicted from the CFT to be marginal, we find that the SRS RG scheme does not exactly reproduce this marginal behavior and instead yields two RG fixed points on the critical line.  Nonetheless, the RG flow can be used to calculate the correlation length critical exponent along the entire line of criticality, and gives a simple closed form expression for this exponent that is in broad agreement with the CFT behavior.  The calculated exponent is exact at the Ising decoupling point (the point in parameter space where the AT model reduces to two uncoupled Ising models) and the relative error in this calculated exponent is comparable to the numerically-intensive MERA simulations performed in Ref.~\cite{Bridgeman2015}.  Specifically, the SRS RG gives a relative error in the prediction of $<10\%$ in the range $\lambda\in [-0.4,0.9]$, and much small error for $|\lambda|\ll 1$, compared with $<5\%$ error in Ref.~\cite{Bridgeman2015}.  The accuracy decreases significantly towards the endpoints of the critical line. While the simplest SRS RG is obtained using the smallest blocking size of $2$ to $1$, we also explore alternate larger blocking schemes, which yield similar results.  

Our paper is structured as follows.  In Sec.~\ref{sec:RGtheory} we introduce the SRS RG scheme in terms of a tree tensor network, focussing on the construction of the basic isometry defining the RG map through properties of the Hamiltonian and its on-site symmetries.  In Sec.~\ref{sec:AT} we apply the SRS RG to the quantum Ashkin-Teller model, constructing this RG map and using it to determine the fixed points and critical behavior of this model along its line of criticality.  We compare the results to the predictions obtained from the orbifold boson CFT, and also explore alternate blockings.  We conclude and discuss these findings in Sec.~\ref{sec:conclusions}.

\section{SRS RG as a concatenated block code}
\label{sec:RGtheory}

\begin{figure}
\centering
\includegraphics{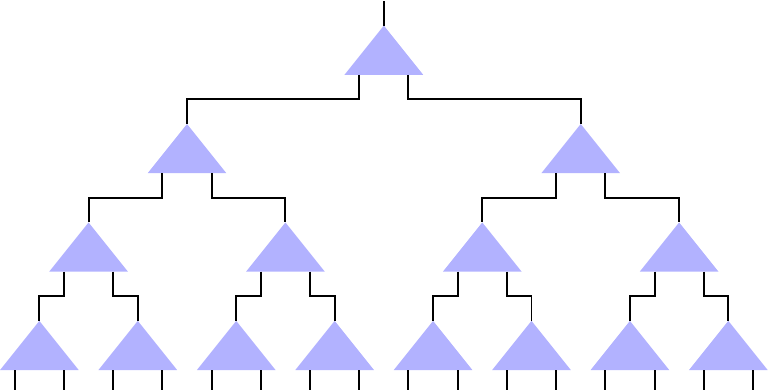}
\caption{Real-space RG as a concatenated block code, presented as a tree tensor network.  Here, the block size is $b=2$, and $4$ layers of concatenation are shown.}
\label{fig:tree}
\end{figure}

The simplest form of real-space renormalization is described by a tree tensor network structure, as in Fig.~\ref{fig:tree}.  From a quantum computing perspective, these tree networks are instances of a concatenated block quantum error correcting code~\cite{Ferris2014}, and the well-studied properties of such codes can provide guidance in choosing an RG scheme with an appropriate structure.

The basic element of an RG scheme of this form is an isometry $W: \mathcal{H}_{d'} \rightarrow \mathcal{H}_d^{\otimes b}$ that maps quantum states of a single $d'$-dimensional spin with Hilbert space $\mathcal{H}_{d'}$ into the Hilbert space of a block of $b$ spins, each with Hilbert space $\mathcal{H}_d$ of dimension $d$.  The isometry condition is $W^\dag W = \mathbb{I}_{d'}$, with $\mathbb{I}_{d'}$ the identity operator on $\mathcal{H}_{d'}$, and this requires $d'\leq d^b$.  We consider only schemes with $d'=d$, meaning that a single $d$-level spin is mapped into a block of $b$ identical spins, but it is straightforward to extend to the general case $d'\neq d$.  

In diagrammatic tensor notation, as in Fig.~\ref{fig:tree}, this isometry is represented by a tensor with one incoming leg and $b$ outgoing legs:
\begin{align}
\begin{array}{l}W \end{array} &= \begin{array}{l}\begin{tikzpicture}[scale=.9,label distance=1mm]
\fill[blue!30] (0.2,0)--(1.8,0)--(1,1)--(0.2,0);
\draw[line width=.2mm] (.5,0)--(.5,-.3);
\draw[line width=.2mm] (.7,0)--(.7,-.3);
\draw[line width=.2mm] (.9,0)--(.9,-.3);
\draw[line width=.2mm] (1.5,0)--(1.5,-.3);
\node at (1.25,-.165){{\scriptsize $\cdots$}};
\draw[line width=.2mm] (1,1)--(1,1.3);
\end{tikzpicture}\end{array}
\label{eqn:BlockingTensor}\end{align}
As an example, a quantum error correcting code that encodes a spin (quantum system) into $b$ spins is specified by an encoding map given by such an isometry $W: \mathcal{H}_d \rightarrow \mathcal{H}_d^{\otimes b}$.  

We require that our blocking preserve the on-site symmetry, as symmetry considerations have been argued to underpin the success of this type of simple real-space RG scheme.  Tensor network methods for quantum many-body systems provides a natural framework to incorporate symmetries~\cite{Singh}.  Let $\mathcal{G}$ be a symmetry group that acts on-site through a unitary action $U_g$, $g \in \mathcal{G}$.  That is, the symmetry acts on a chain of $N$ spins as $U_g^{\otimes N}$, and the Hamiltonian $H_N$ for $N$ spins satisfies $[H_N, U_g^{\otimes N}]=0$ for all $g\in \mathcal{G}$.  Our block encoding then preserves the on-site symmetry of the group $\mathcal{G}$ if it intertwines the representation $U^{\otimes b}_g$ on $b$ input spins with a representation $V_g$ on the output spin, i.e.,
\begin{equation}
	U_g^{\otimes b} W = W V_g \,, \quad \forall g \in \mathcal{G}\,.
\end{equation}
This can be expressed as a tensor relation (shown here for a block size of $b=2$) as
\begin{align}
\begin{array}{l}\begin{tikzpicture}[scale=.9,label distance=1mm]
\fill[blue!30] (0.2,0)--(1.8,0)--(1,1)--(0.2,0);
\draw[line width=.2mm] (.5,0)--(.5,-.6);
\draw[line width=.2mm] (1.5,0)--(1.5,-.6);
\draw[line width=.2mm] (1,1)--(1,1.6);
\node[circle,fill=red!30,label=below:$U_g$,align=center,anchor=north](node1) at (.5,-.15){};
\node[circle,fill=red!30,label=below:$U_g$,align=center,anchor=north](node1) at (1.5,-.15){};
\node[circle,label=above:\textcolor{white}{$V_g$},align=center,anchor=south](node1) at (1.5,1.15){};
\end{tikzpicture}\end{array}&=\begin{array}{l}\begin{tikzpicture}[scale=.9,label distance=1mm]
\fill[blue!30] (0.2,0)--(1.8,0)--(1,1)--(0.2,0);
\draw[line width=.2mm] (.5,0)--(.5,-.6);
\draw[line width=.2mm] (1.5,0)--(1.5,-.6);
\draw[line width=.2mm] (1,1)--(1,1.6);
\node[circle,label=below:\textcolor{white}{$U_g$},align=center,anchor=north](node1) at (.5,-.15){};
\node[circle,label=below:\textcolor{white}{$U_g$},align=center,anchor=north](node1) at (1.5,-.15){};
\node[circle,fill=red!30,label=above:$V_g$,align=center,anchor=south](node1) at (1,1.15){};
\end{tikzpicture}\end{array}&\begin{array}{l}\forall\ g\in\mathcal{G}\,.\end{array}\label{eqn:SymmetryCondition}\end{align}
Again, we can look to quantum error correcting codes for examples.  A quantum gate $T$ is said to act \emph{transversally} on a code if the action $T^{\otimes b}$ on all physical spins in the block is equivalent to the encoded action of $T$.  Thus, any code with transversal action of all gates $U_g$, $g\in \mathcal{G}$ provides a blocking that satisfies Eq.~(\ref{eqn:SymmetryCondition}) where $V$ is the original on-site representation $U$.

A real-space RG scheme that has the form of a tree tensor network is now entirely determined by identifying an isometry $W$.  We now use the structure of the system's Hamiltonian to identify the specific choice of this isometry.  We seek to separate terms in the total system Hamiltonian $H$ into groups $H^I_{\rm in}$ consisting of terms that act within a block $I$, and the remaining terms $H_{\rm rest}$, such that
\begin{equation}
	H = H_{\rm rest} + \sum_{I \in {\rm blocks}} H^I_{\rm in} \,.
\end{equation}
For simplicity, consider the translationally-invariant case where all blocks are identical, meaning $H^I_{\rm in} \equiv H_{\rm in}$ for all $I$.  There is considerable freedom in this decomposition, as not all terms with support in a block must necessarily be included in $H_{\rm in}$.  

Following the approach of Refs.~\cite{Miyazaki2011,Kubica}, we place some specific constraints on the choice of $H_{\rm in}$ that will in turn yield a suitable RG map.  First, note that any choice of block Hamiltonian $H_{\rm in}$ defines a linear projection map $E_0 :\mathcal{H}_d^{\otimes b} \rightarrow \mathcal{H}_{d_0}$, where $\mathcal{H}_{d_0}$ is the ground space of $H_{\rm in}$ and $d_0$ its dimension.  If $d_0>1$, then the ground state space of $H_{\rm in}$ is degenerate; from a code perspective, we say that such an  $H_{\rm in}$ with a degenerate ground space defines a code $\mathcal{H}_{d_0} \subset \mathcal{H}_d^{\otimes b}$.  For our RG scheme, we seek to choose our block Hamiltonian $H_{\rm in}$ with a degenerate ground space of dimension $d_0=d$, with projection $E_0$ onto this degenerate ground space.  By defining $W = E_0^\dagger$, we obtain an isometry as required for our RG map.  

In addition, we require that $H_{\rm in}$ is invariant under the symmetry, $[H_{\rm in},U_g^{\otimes b}]=0$ for all $g \in \mathcal{G}$.  This condition ensures that the ground space $\mathcal{H}_{d_0}$ of $H_{\rm in}$ carries a representation $V$ of $\mathcal{G}$.   The resulting isometry $W = E_0^\dagger$ will then transform under $\mathcal{G}$ according to Eq.~(\ref{eqn:SymmetryCondition}) as required.  A natural restriction is to the \emph{scale-invariant} case where $V = U$, the original on-site representation of $\mathcal{G}$.  In this case, the degenerate ground space of $H_{\rm in}$ acting on $b$ spins has the dimension of a single spin, and the symmetry group $\mathcal{G}$ acts on this degenerate ground space exactly as it acts on a single spin, i.e., transversally.

One round of blocking for our RG map renormalizes the Hamiltonian as $H \rightarrow H'$, where
\begin{equation}
	H' = \bigl( \otimes_I W^\dagger_I \bigr) H \bigl( \otimes_I W_I \bigr) \,.
\end{equation}
As by construction we have that $W_I^\dagger H^I_{\rm in} W_I = c\mathbb{I}_{d}$ for $\mathbb{I}_d$ the identity operator on the renormalized spin from block $I$ and $c$ a constant, the renormalized Hamiltonian is determined by
\begin{equation}
	H' = \bigl( \otimes_I W^\dagger_I \bigr) H_{\rm rest} \bigl( \otimes_I W_I \bigr) \,,
\end{equation}
up to an irrelevant additive constant, which is now an operator acting on the renormalized spins.

A real-space RG scheme such as this has been applied to the transverse field Ising model by Miyazaki \textit{et al.}~\cite{Miyazaki2011}, based on the $\mathbb{Z}_2$ spin-flip symmetry of the Ising model, where the resulting $2$-to-$1$ RG produces the exact expression for the critical exponent $\nu$ that describes the scaling of the Ising model correlation length.  This method is also surprisingly accurate when applied to 2D quantum Ising models on a variety of lattices~\cite{Miyazaki2011,Kubica}.  

\section{Real-space renormalization of the quantum Ashkin-Teller model}
\label{sec:AT}

In this section, we will apply the SRS RG method described above to the Ashkin-Teller model.  Note that the critical behavior of this model has previously been investigated using real-space RG techniques in Ref.~\cite{Igloi}, wherein both a standard blocking method and a related (but distinct) symmetry-respecting method were used.

\subsection{The quantum Ashkin-Teller model}
\label{sec:ATintro}

The quantum Ashkin-Teller (AT) model~\cite{Solyom1981,Ashkin1943} is defined by the Hamiltonian
  \begin{align}
      H_{\rm{AT}}=&-J \sum_j \bigl( Z_{j,1} Z_{j+1,1} +  Z_{j,2} Z_{j+1,2} \bigr. \nonumber \\
      & \qquad\qquad \bigl. + \lambda Z_{j,1} Z_{j+1,1} Z_{j,2} Z_{j+1,2} \bigr) \nonumber \\
      &  -h\sum_j \bigl(X_{j,1} + X_{j,2} + \lambda X_{j,1}X_{j,2} \bigr)\,,
      \label{eq:ATHam}
  \end{align}
with $J$, $h$, and $\lambda$ real-valued coupling coefficients.  This model can be viewed as a pair of transverse-field Ising chains coupled by two- and four- spin terms. In particular, when $\lambda=0$, we have a decoupled pair of quantum transverse-field Ising chains.

The AT model possesses an on-site $D_4$ symmetry, defined as follows.  Define the operators
  \begin{equation}
    S_1 = \bigotimes_j X_{j,1} \,, \qquad S_2 = \bigotimes_j X_{j,2}	\,.
  \end{equation}
These operators commute with the Hamiltonian, and generate a $\mathbb{Z}_2 \times \mathbb{Z}_2$ symmetry.  The Hamiltonian is also invariant under the symmetry that swaps the two chains, i.e., $X_{j,1}\leftrightarrow X_{j,2}$ and $Z_{j,1}\leftrightarrow Z_{j,2}$.  Together, these symmetries form the symmetry group $D_4$.
(We note that this model also possesses a \emph{non-local} symmetry, given by applying the self-duality map of the Ising model to each of the two chains.  This self-duality ensures that, for a fixed value of $\lambda$, if there is a single phase transition it must be at $h = J$.)

The model is critical along the line $\beta\equiv h/J=1$ and $-1/\sqrt{2} < \lambda\leq 1$, and the critical indices vary continuously along this line.  Based on finite size simulation~\cite{Baake1987}, CFT arguments~\cite{Yang1987}, and real-space RG based on MERA~\cite{Bridgeman2015}, the critical line of the AT model has been identified as the orbifold boson with radius $R_O$ given by
  \begin{align}
    R_O^2=\frac{\pi}{2}[\arccos(-\lambda)]^{-1}\,, \label{eqn:Ro}
  \end{align}
  for $\lambda \in (-1/\sqrt{2},1]$.  
In addition, the AT model for $\beta=1$ for any $\lambda$ can be transformed into the six-vertex model~\cite{Kohmoto}, and so is exactly solvable using Bethe Ansatz.  These relations along the critical line to the six-vertex model and the orbifold boson CFT allows us to obtain analytic predictions for the critical behavior of this model along the line $\beta=1$~\cite{Kohmoto,DiFrancesco1997}, which is governed entirely by the radius $R_O^2$ of the orbifold boson.  

In particular, consider the correlation length $\xi$ defined as for the Ising model $(\langle Z_{i,1} Z_{j,1}\rangle - \langle Z_{i,1} \rangle^2) \sim e^{-(j-i)/\xi}$, where we have arbitrarily chosen the first Ising chain but symmetry ensures that it is the same for both chains.  With $\beta=h/J$ as a tuning parameter and $\beta_0=1$ at the critical point, the correlation length near the critical point is governed by $\xi = (\beta-\beta_0)^{-\nucrit}$, with $\nucrit$ the \emph{correlation length critical exponent}.  This critical exponent $\nucrit$ %, which is related to the critical exponent $\alpha$ through the standard relation $\nu d = (2-\alpha)$ with $d=2$ in this case, 
    is expected to vary continuously along the line of criticality.  From Ref.~\cite{Kohmoto}, we have a theoretically predicted value of $\nucrit$ given in terms of the orbifold boson radius $R_O$ as $\nucrit = 1/(2-R_O^2)$ and therefore
    %$\alpha=2-2/(2-R_O^2)$.  The theoretically predicted value $\nucrit$ of the Ashkin-Teller model along the critical line is therefore
\begin{equation}
  \label{eq:xT}
  \nucrit = \frac{1}{2-\frac{\pi}{2}[\arccos(-\lambda)]^{-1}} \,.
\end{equation}
Based on the success of predicting this critical exponent for the quantum transverse-field Ising model (where $\nucrit(\lambda=0)=1$) using the SRS RG, we now aim to calculate a prediction for this exponent in the quantum AT model.  

\subsection{Symmetries and block Hamiltonians}
\label{sec:symmetries}

Here we describe an SRS RG map for the Ashkin-Teller model as a concatenated block code that respects the $D_4$ symmetry of the model.  This on-site $D_4$ symmetry acts on a pair of spins (labelled by $(j,1)$ and $(j,2)$ in the Hamiltonian of Eq.~(\ref{eq:ATHam})), that is, a `site' consists of a pair of spins and is labelled by $j$.

The simplest blocking, which we consider first, is a $2$-to-$1$ blocking, i.e., that maps $4$ spins (two sites) to $2$ spins (one site).  We therefore seek an isometry $W: \mathcal{H}_4 \rightarrow \mathcal{H}_4^{\otimes 2}$ that encodes a $4$-dimensional quantum spin into a pair of $4$-dimensional spins.  To identify an appropriate isometry, we select terms in the Hamiltonian~(\ref{eq:ATHam}) to define a block Hamiltonian $H_{\rm in}$ on a pair of sites, possessing a $4$-dimensional degenerate ground state on which the $D_4$ symmetry acts transversally. 

As a starting point based on the successful solution of this form for the quantum Ising model with its $\mathbb{Z}_2$ symmetry as in Refs.~\cite{Miyazaki2011,Kubica}, one choice of a block Hamiltonian satisfying these conditions is
\begin{multline}
  H_{\rm in}=-J \bigl(Z_{j,1} Z_{j+1,1} + Z_{j,2} Z_{j+1,2} \\
  \qquad + \lambda Z_{j,1} Z_{j+1,1} Z_{j,2} Z_{j+1,2} \bigr) \\ -h \bigl(X_{j,1} + X_{j,2} +\lambda X_{j,1} X_{j,2} \bigr) \,,
\end{multline}
expressed here as acting on sites $j$ and $j+1$ within a block (e.g., by choosing $j$ odd).  This Hamiltonian has a $4$-dimensional degenerate ground state on which the $D_4$ symmetry acts transversally.  These properties are seen most easily from the perspective of the Hamiltonian code defined by this block Hamiltonian's ground space.  Defining the logical operators for a $2$-spin phase-flip redundancy code
\begin{align}
  \bar{X}_1&=X_{j,1} X_{j+1,1} \,, &\quad
  \bar{Z}_1&=Z_{j+1,1} \,, \\
  \bar{X}_2&=X_{j,2} X_{j+1,2} \,, &\quad
  \bar{Z}_2&=Z_{j+1,2} \,,
  \label{eq:ATbar}
\end{align}
we see that these operators commute with $H_{\rm in}$, acting as a pair of renormalized Pauli operators on the degenerate ground space.

\subsection{Solving the block Hamiltonian}

To find an explicit expression for the isometry $W$ defined by an embedding of the encoded site into the ground space, we need to solve $H_{\rm in}$.  For convenience, we change to a new set of operators
\begin{align}
\tilde{X}_1&= X_{j,1}\,, &\quad
\tilde{Z}_1&= Z_{j,1} Z_{j+1,1}\,, \\
\tilde{X}_2&= X_{j,2}\,, &\quad
\tilde{Z}_2&= Z_{j,2} Z_{j+1,2}\,. 
\label{eq:ATtilde}
\end{align}
These operators commute with the logical operators of Eq.~(\ref{eq:ATbar}), and we can express $H_{\rm in}$ entirely in terms of these operators as
\begin{multline}
  J^{-1} \tilde{H}_{\rm in}=- \tilde{Z}_1 -\tilde{Z}_2 -\beta
\tilde{X}_1-\beta \tilde{X}_2 \\ -\lambda \beta
\tilde{X}_1\tilde{X}_2 -\lambda  \tilde{Z}_1\tilde{Z}_2 \,,
\label{eq:Hint}
\end{multline}
where we recall that $\beta \equiv h/J$.  We note while $H_{\rm in}$ is 4-fold degenerate on the 4 spins in the block, the Hamiltonian $\tilde{H}_{\rm in}$ is nondegenerate in terms of the operators (\ref{eq:ATtilde}).  This is because $H_{\rm in}$ is supported on a single pair of effective spins, as follows.  It is most natural to view the space $\mathcal{H}_4^{\otimes 2}$ of the two sites as possessing a virtual tensor product structure $\tilde{\mathcal{H}}_4 \otimes \bar{\mathcal{H}}_4$, where $\tilde{\mathcal{H}}_4$ corresponds to the support of the operators (\ref{eq:ATtilde}) and $\bar{\mathcal{H}}_4$ the support of the logical operators (\ref{eq:ATbar}).  Thus, the block Hamiltonian takes the form $H_{\rm in} = \tilde{H}_{\rm in} \otimes \bar{I}_4$, where having $\tilde{H}_{\rm in}$ nondegenerate ensures that $H_{\rm in}$ is $4$-fold degenerate.  

This Hamiltonian can be solved analytically.  It has a particularly simple solution along the line $\beta=1$, where the ground-state energy is given by 
\begin{equation}
  E_g(\beta=1,\lambda)/J =-\lambda -\sqrt{\lambda^2+8}\,,
\end{equation}
and the ground state $| \tilde{g}_{\beta=1,\lambda} \rangle$ can be expressed as 
\begin{equation}
  \label{eq:criticallinesolution}
| \tilde{g}_{\beta=1,\lambda} \rangle = \cos(\theta/2)| \phi \rangle |\phi  \rangle+\sin(\theta/2)| \phi^\perp \rangle |\phi^\perp \rangle\,.
\end{equation}
Here, we have defined the state $|\phi\rangle$ as the $+1$ eigenstate of the operator $(\tilde{X}+\tilde{Z})/\sqrt{2}$, $|\phi^\perp\rangle$ is the orthogonal spin state with eigenvalue $-1$, and  $\tan\theta = \lambda/(2\sqrt{2})$. 

The general solution for the eigenvalues and eigenstates of the block Hamiltonian $H_{\rm in}$ of Eq.~(\ref{eq:Hint}) are closed-form cubic expressions (not shown).  The ground state $|\tilde{g}_{\beta,\lambda}\rangle$ is nondegenerate for $\lambda \in (-1,1]$, and there are no crossings. 

The isometry $W$ then takes the simple form 
\begin{equation}
  W = |\tilde{g}_{\beta,\lambda}\rangle \otimes \bar{I}_4\,,
\end{equation}
in terms of the virtual tensor product structure $\tilde{\mathcal{H}}_4 \otimes \bar{\mathcal{H}}_4$ given by the operators (\ref{eq:ATtilde}) and (\ref{eq:ATbar}).

\subsection{Renormalizing the Hamiltonian}

Applying one round of blocking for our RG map results in a renormalized Hamiltonian given by
\begin{equation}
	H' = \bigl( \otimes_I W^\dagger_I \bigr) H \bigl( \otimes_I W_I \bigr) \,,
\end{equation}
which, in terms of our new operators and the solution $|\tilde{g}_{\beta,\lambda}\rangle$ for the ground space of $H_{\rm in}$, takes the form
\begin{align}
H'=-J \sum_{I} &\Bigl[ \langle \tilde{g}_{\beta,\lambda} | \tilde{Z}_1 | \tilde{g}_{\beta,\lambda} \rangle \bar{Z}_{1,I} \bar{Z}_{1,I+1} \nonumber\\
& + \langle \tilde{g}_{\beta,\lambda}|\tilde{Z}_2 | \tilde{g}_{\beta,\lambda} \rangle {\bar{Z}_{2,I}}{\bar{Z}_{2,I+1}} \nonumber\\
& +\lambda \langle \tilde{g}_{\beta,\lambda} |\tilde{Z}_1\tilde{Z}_2 | \tilde{g}_{\beta,\lambda} \rangle \bar{Z}_{1,I}\bar{Z}_{1,I+1} {\bar{Z}_{2,I}}{\bar{Z}_{2,I+1}} \Bigr] \nonumber \\
-h \sum_I &\Bigl[ \langle \tilde{g}_{\beta,\lambda} |\tilde{X}_1 | \tilde{g}_{\beta,\lambda} \rangle{\bar{X}_{1,I}} + \langle \tilde{g}_{\beta,\lambda} |\tilde{X}_2 | \tilde{g}_{\beta,\lambda} \rangle {\bar{X}_{2,I}} \nonumber \\
&+ \lambda \langle \tilde{g}_{\beta,\lambda} |\tilde{X}_1\tilde{X}_2 | \tilde{g}_{\beta,\lambda} \rangle \bar{X}_{1,I}  {\bar{X}_{2,I}} \Bigr] \,,
 \end{align}
where $I$ denotes the $I$th site in the renormalised chain (a block of two sites in the original chain).

Because the Hamiltonian $H_{\rm in}$ has a symmetry swapping $1\leftrightarrow 2$, we have identities $\langle \tilde{g}_{\beta,\lambda} |\tilde{{Z}}_1 | \tilde{g}_{\beta,\lambda} \rangle = \langle \tilde{g}_{\beta,\lambda} |\tilde{Z}_2 | \tilde{g}_{\beta,\lambda} \rangle$, etc.  Therefore, the renormalized Hamiltonian can be expressed as
\begin{multline}
H'
=- J' \bar{Z}_{1,I} \bar{Z}_{1,I+1} - J' {\bar{Z}_{2,I}}{\bar{Z}_{2,I+1}} -h' {\bar{X}_{1,I}} -h' {\bar{X}_{2,I}}  \\
-\lambda_{X}' h' \bar{X}_{1,I}  {\bar{X}_{2,I}} 
 -\lambda_{Z}' J' \bar{Z}_{1,I} \bar{Z}_{1,I+1}{\bar{Z}_{2,I}}{\bar{Z}_{2,I+1}}
\end{multline}
where the coefficients are given by
\begin{align}
	J'/J &= \langle \tilde{g}_{\beta,\lambda} |\tilde{Z}_1 | \tilde{g}_{\beta,\lambda} \rangle \,, \label{eq:j}\\
	h'/h &= \langle \tilde{g}_{\beta,\lambda} |\tilde{X}_1 | \tilde{g}_{\beta,\lambda} \rangle \,, \label{eq:h}\\
	\lambda'_{X}/\lambda_X &= (h/h') \langle \tilde{g}_{\beta,\lambda} |\tilde{X}_1 \tilde{X}_2| \tilde{g}_{\beta,\lambda} \rangle \nonumber \\
	 &= \langle \tilde{g}_{\beta,\lambda} |\tilde{X}_1 \tilde{X}_2| \tilde{g}_{\beta,\lambda} \rangle/ \langle \tilde{g}_{\beta,\lambda} |\tilde{X}_1 | \tilde{g}_{\beta,\lambda} \rangle \,, \label{eq:l1}\\
	\lambda'_{Z}/\lambda_Z &= (J/J') \langle \tilde{g}_{\beta,\lambda} |\tilde{Z}_1 \tilde{Z}_2| \tilde{g}_{\beta,\lambda} \rangle \nonumber \\
	&= \langle \tilde{g}_{\beta,\lambda} |\tilde{Z}_1 \tilde{Z}_2| \tilde{g}_{\beta,\lambda} \rangle/ \langle \tilde{g}_{\beta,\lambda} |\tilde{Z}_1 | \tilde{g}_{\beta,\lambda} \rangle \,. \label{eq:l2} 
\end{align}
We note that the two terms with $\lambda$ in the Hamiltonian do not rescale the same way, and so we have defined $\lambda_X h$ as the coefficient of the $XX$ term and $\lambda_Z J$ as the coefficient of the $ZZZZ$ term.  The symmetry of the AT model that relates these two terms is the self-duality symmetry of the Ising model, which is a nonlocal symmetry that is not enforced in the SRS RG map.

\subsubsection{Identifying the critical line}

From Eqs.~(\ref{eq:j}-\ref{eq:h}) and our solution $|\tilde{g}_{\beta,\lambda}\rangle$, we find a line of fixed points for the $h$ and $J$ coefficients along the line $\beta=h/J=1$.  That is, from the flow of these coefficients we recover the known critical line.  

We note, however, that the coefficients $\lambda_{X,Z}$ are not fixed points along this entire line for the SRS RG map.  We can use our analytical solution of Eq.~(\ref{eq:criticallinesolution}) along the critical line $\beta=1$ to evaluate
\begin{align}
\langle \tilde{g}_{\beta,\lambda} |\tilde{Z}_1 | \tilde{g}_{\beta,\lambda} \rangle  &=\frac{2}{\sqrt{\lambda^2+8}} \\
\langle \tilde{g}_{\beta,\lambda} |\tilde{X}_1 | \tilde{g}_{\beta,\lambda} \rangle & =\frac{2}{\sqrt{\lambda^2+8}} \\
\langle \tilde{g}_{\beta,\lambda} |\tilde{Z}_1 \tilde{Z}_2| \tilde{g}_{\beta,\lambda} \rangle  &= \frac{1}{2} \Bigl( 1 + \frac{\lambda}{\sqrt{\lambda^2+8}} \Bigr) \\
\langle \tilde{g}_{\beta,\lambda} |\tilde{X}_1 \tilde{X}_2| \tilde{g}_{\beta,\lambda} \rangle &= \frac{1}{2} \Bigl( 1 + \frac{\lambda}{\sqrt{\lambda^2+8}} \Bigr) \,.
\end{align}
We then find that the coefficients rescale along the $\beta=1$ line according to
\begin{align}
h'&= \frac{2}{\sqrt{\lambda^2+8}}h 
\label{eq:hexact} \\
J'&=\frac{2}{\sqrt{\lambda^2+8}}J \label{eq:jexact} \\
\lambda'&= \frac{\lambda}{4}(\lambda+\sqrt{\lambda^2+8}) \,.\label{eq:lamexact}
\end{align}
Note that the SRS RG map has two fixed points, $\lambda_*=0$ and $\lambda_* =1$.  Note these fixed points are the same as those found in the related real-space RG scheme of Ref.~\cite{Igloi}.

When the SRS RG map is evaluated along the critical line $\beta=1$, the two parameters $\lambda_X$ and $\lambda_Z$ flow in the same way, i.e., the expressions (\ref{eq:l1}) and (\ref{eq:l2}) are identical at $\beta=1$.  (This is not true for general $\beta$.)  As a result, when considering the critical line we can dispense with the need for two different coefficients and return to using the single coefficient $\lambda$ to describe critical behavior.

\subsubsection{Calculating critical exponents}

We now use the SRS RG map to determine the behavior along along the critical line $\beta=1$.  Although the SRS RG map results in a flow along this critical line to two fixed points ($\lambda_*=0$ and $\lambda_* =1$), we show that we can use a linearised solution around any point along the model's critical line to estimate the correlation length critical exponent of the AT model, which turns out to be independent of the SRS RG map's flow along the critical line.

About any point along the critical line $(\beta=1, \lambda \in (-1/\sqrt{2},1])$, the behavior is governed by the Jacobian of a two-parameter transformation $(\beta,\lambda) \rightarrow (\beta',\lambda')$.  The closed-form solution for the matrix elements of this Jacobian using Eqs.~(\ref{eq:j}-\ref{eq:l2}) and $|\tilde{g}_{\beta,\lambda}\rangle$ is 
\begin{align}
	\left. \frac{\partial \beta'}{\partial \beta}\right|_{\beta=1} &= 2 + \frac{\lambda}{4}(-\lambda+\sqrt{\lambda^2+8}) \,, \\
	\left.\frac{\partial \lambda'}{\partial \beta}\right|_{\beta=1} &= 0 \,, \\
	\left.\frac{\partial \beta'}{\partial \lambda}\right|_{\beta=1} &= \frac{\lambda}{8}(1-\lambda^2)(-\lambda+\sqrt{\lambda^2+8}) \,, \\
	\left.\frac{\partial \lambda'}{\partial \lambda}\right|_{\beta=1} &=  \frac{2+\frac{1}{2}\lambda(\lambda+\sqrt{\lambda^2+8})}{\sqrt{\lambda^2+8}} \,.
\end{align}
The eigenvalues of the Jacobian, which can be expressed in the form $b^{y_{1,2}}$ where $b$ is the block decimation size ($b=2$ in this example, being a mapping of $4$ spins to $2$), and $\nu_{1,2}$ are functions of $\lambda \in (-1/\sqrt{2},1]$.  As the Jacobian is a triangular matrix, the eigenvalues are simply the diagonal entries, i.e.,
\begin{align}
	y_1(\lambda) &= \log_2 \Bigl[2 + \frac{\lambda}{4}(-\lambda+\sqrt{\lambda^2+8}) \Bigr]\,, \\ 
	y_2(\lambda) &= \log_2 \Bigl[ \frac{2+\frac{1}{2}\lambda(\lambda+\sqrt{\lambda^2+8})}{\sqrt{\lambda^2+8}} \Bigr] \,.
\end{align}

\begin{figure}
\includegraphics{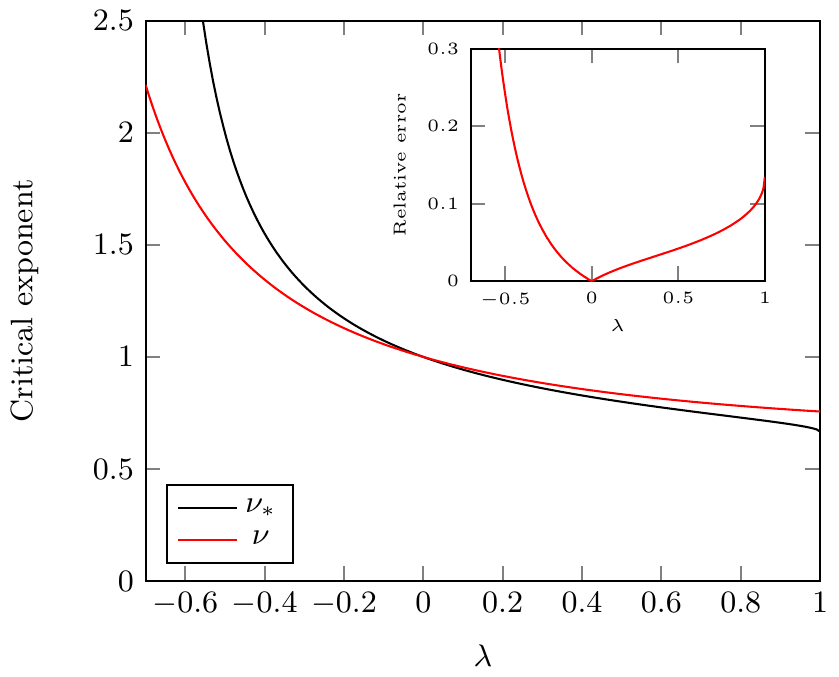}
\caption{Comparison of the critical index $\nucrit$ of the energy operator (black) with the critical exponent $\nupred$ obtained from our real-space RG map (red) as functions of the Hamiltonian parameter $\lambda$, using the smallest blocking size $b=2$.  Inset:  Relative error $|\nupred-\nucrit|/\nucrit$ of the predicted critical exponent $\nupred$ obtained from our real-space RG map to the critical index $\nucrit$.}
\label{fig:AT_42_analytic}
\end{figure}

The first of these eigenvalues $y_1$ associated with $\partial \beta'/\partial \beta$ gives the estimate of the correlation length critical exponent as $\nupred = y_1^{-1}$.  As discussed in Sec.~\ref{sec:ATintro}, the value of this critical exponent is predicted to be $\nucrit$ as given by Eq.~(\ref{eq:xT}).  Comparing $\nupred(\lambda)$ and $\nucrit(\lambda)$ in Fig.~\ref{fig:AT_42_analytic}, we see broad agreement, including exact agreement at $\lambda=0$, where $\nupred(0) = \nucrit(0) = 1$ but decreasing accuracy towards the limits $\lambda=-1/\sqrt{2}$ and $\lambda=1$.  For comparison, at $\lambda=1$ we have $\nucrit(1) = 2/3$ compared with $\nupred(1) = (\log_2 5 - 1)^{-1} \simeq 0.75$.  (Note that, at $\lambda=1$, the AT model is equivalent to the 4-state Potts model.)  At $\lambda= -1/\sqrt{2}$, we have $\nucrit(-1/\sqrt{2}) \to \infty$ compared with $\nupred(-1/\sqrt{2}) \simeq 2.25$.

The second eigenvalue $y_2$ is associated with the flow along the critical line that is predicted by the SRS RG map, with fixed points $\lambda=0$ (with $y_2 = -1/2$ at this point) and $\lambda =1$ (with $y_2 = (2-\log_2 3)$).  We contrast this flow to predicted behavior of the AT model, where the corresponding scaling field is marginal.

\subsection{Alternative blockings}

Our analysis so far has restricted to blocking of $4$ spins to $2$ (i.e., $b=2$).  In this section, we explore alternate blocking choices for the SRS RG of the quantum AT model.

For an alternative blocking choice $b$, we seek an isometry $W_b: \mathcal{H}_4 \rightarrow \mathcal{H}_4^{\otimes b}$ that encodes a 4-dimensional quantum spin into $b$ such spins.  Following Sec.~\ref{sec:symmetries}, a natural generalization of the $b=2$ case is to use the phase-flip redundancy code with logical operators
\begin{align}
  \bar{X}_1&= X_{j,1} X_{j+1,1} \cdots X_{j+b-1,1}  \,, &\quad
  \bar{Z}_1&=Z_{j+k_0,1} \,, \\
  \bar{X}_2&=X_{j,2} X_{j+1,2} \cdots X_{j+b-1,2} \,, &\quad
  \bar{Z}_2&=Z_{j+k_0,2} \,,
  \label{eq:ATbar}
\end{align}
where $k_0\in \{0,\ldots,b-1\}$ is an integer specifying a preferred site in the block.  These operators commute with the following choice of block Hamiltonian
\begin{multline}
  H_{\rm in}=-J \sum_{k=0}^{b-1} \bigl(Z_{j+k,1} Z_{j+k+1,1} + Z_{j+k,2} Z_{j+k+1,2} \\
  \qquad + \lambda Z_{j+k,1} Z_{j+k+1,1} Z_{j+k,2} Z_{j+k+1,2} \bigr) \\ -h \sum_{k\neq k_0} \bigl(X_{j+k,1} + X_{j+k,2} +\lambda X_{j+k,1} X_{j+k,2} \bigr) \,.
\end{multline}
Note that this block Hamiltonian includes all terms in the quantum AT Hamiltonian with support entirely within the block, \emph{except} for the $X$- and $XX$-type terms on the preferred site $k_0$.  This choice of block Hamiltonian has a $4$-fold degenerate ground space, and therefore defines an isometry $W$ with the desired properties of our real-space RG map.

\begin{figure}
\includegraphics{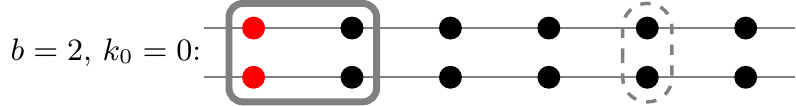}
\includegraphics{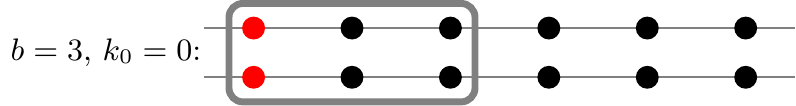}
\includegraphics{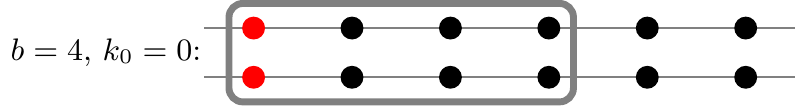}
\includegraphics{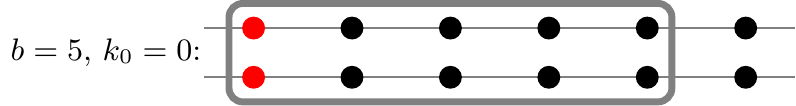}
\includegraphics{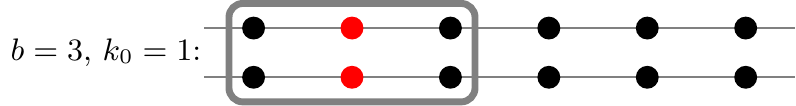}
\includegraphics{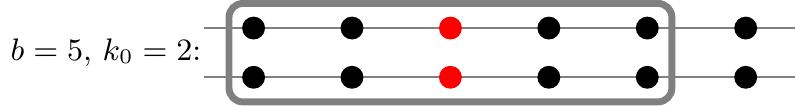}
\caption{Alternative choices of blocking.  Within each block consisting of $b$ sites, one site (pair of spins, such as shown encircled by dashed line), shown in red, is a preferred site for the encoding map.}
\label{fig:Blockings}
\end{figure}

We consider two choices of preferred site, closely following Ref.~\cite{Kubica}:  $k_0=0$, corresponding to the preferred site at the edge of a block, and $k_0=(b-1)/2$ for odd $b$, corresponding to the middle site of a block.  For $b>2$, we solve $H_{\rm in}$ and determine the isometry $W$ numerically, for a range of block sizes as illustrated in Fig.~\ref{fig:Blockings}.

\begin{figure}
\includegraphics{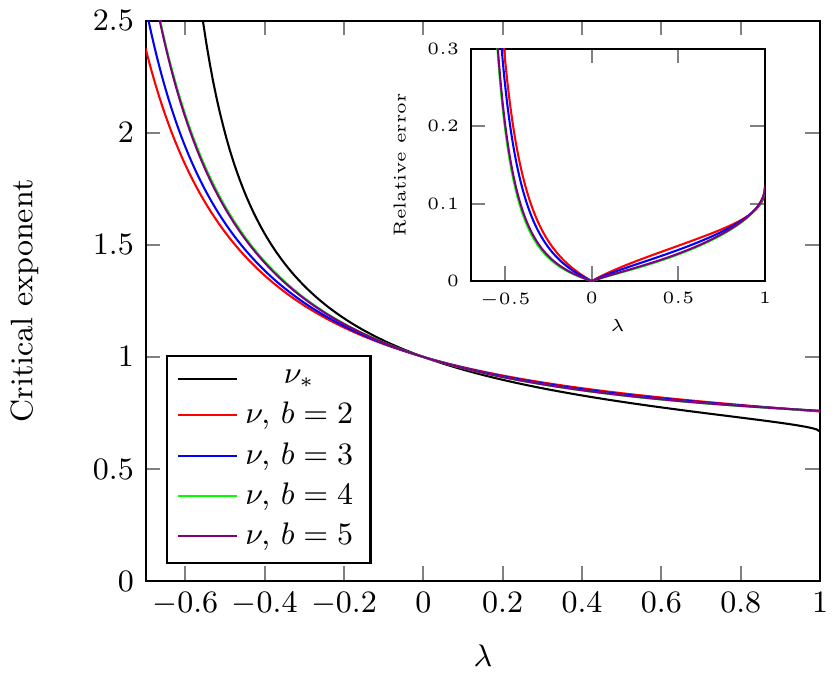}
\caption{Critical exponent $\nupred$ obtained from our real-space RG map using various blocking sizes $b$, all with the preferred site $k_0=0$ on the edge of the block, as illustrated in Fig.~\ref{fig:Blockings}.  For comparison, the exact value $\nucrit$ is shown (black line).   Inset:  Relative error $|\nupred-\nucrit|/\nucrit$ of the predicted critical exponent $\nupred$ obtained for each choice.}
\label{fig:AT_plot_blocks}
\end{figure}

\begin{figure}
\includegraphics{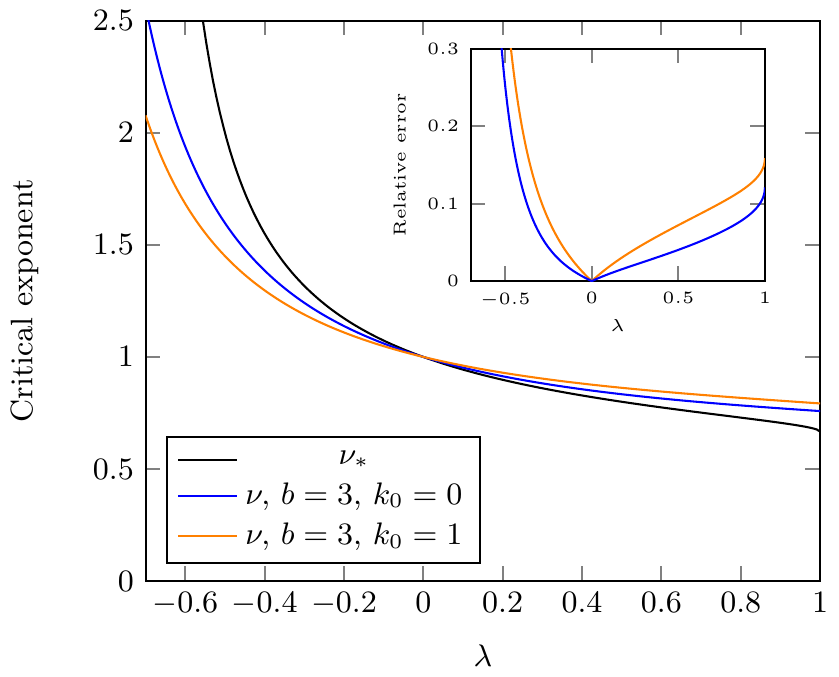}
\caption{Critical exponent $\nupred$ obtained from our real-space RG map using a blocking size $b=3$ with a symmetric (orange) vs asymmetric (blue) choice of internal Hamiltonian terms as illustrated in Fig.~\ref{fig:Blockings}.  For comparison, the exact value $\nucrit$ is shown (black line).   Inset:  Relative error $|\nupred-\nucrit|/\nucrit$ of the predicted critical exponent $\nupred$ obtained for each choice. }
\label{fig:AT_plot_sym}
\end{figure}

The predicted critical exponent $\nupred$ is found to be largely independent of the size of blocking $b$, as shown in Fig.~\ref{fig:AT_plot_blocks}.  All choices of blocking recover the exact result $\nucrit=1$ at $\lambda=0$, as in Ref.~\cite{Kubica}.  For $\lambda>0$, all blockings with preferred site $k_0=0$ at the edge give nearly indistinguishable predictions, where for $\lambda<0$ the accuracy increases with increased blocking size but only marginally, with negligible improvement observed beyond $b=4$.  For $b=3$, we show a comparison of the choice of preferred spin $k_0=0$ (edge of the block) and $k_0=1$ (center of block), shown in Fig.~\ref{fig:AT_plot_sym}.  Again, both cases recover $\nucrit=1$ at $\lambda=0$ exactly.  The choice $k_0=0$ gives more accurate results for $\lambda>0$.  For $\lambda<0$, we see the largest qualitative differences of all blocking choices but still have broad agreement.  For $b=5$ we have performed a similar comparison of $k_0=0$ and $k_0=2$ (not shown), where we find similar but less pronounced differences.

\section{Conclusions}
\label{sec:conclusions}

We have presented a generalization of the simple real-space RG scheme proposed in Ref.~\cite{Miyazaki2011} and further explored in Ref.~\cite{Kubica}, based on a blocking isometry that respects the symmetry of the model.  We have used this SRS RG to investigate the critical behavior of the quantum Ashkin-Teller model, which possesses a line of criticality along which the critical exponents vary continuously, and which reduces to decoupled Ising models at one point on this line.  The SRS RG map was used to extract estimates of the correlation length critical exponent $\nupred$ as a function on this line of criticality for various blocking schemes, including a closed-form expression for the simplest choice of blocking.

The SRS RG identified the line of criticality in this model, and the resulting estimate for the correlation length critical exponent $\nupred$ varied continuously along the critical line.  Compared with the exact quantity $\nucrit$, it gave a precise value at the Ising decoupling point and reproduced the broad features of its functional form along this line as a simple analytical expression with the smallest block size.  In a range of $\lambda$ around the decoupled Ising point $\lambda=0$, the accuracy of the SRS RG is good, with a relative error in the prediction of $<10\%$ in the range $\lambda\in [-0.4,0.9]$.  We compare this accuracy to that obtained by the sophisticated and numerically-intensive MERA simulations performed in Ref.~\cite{Bridgeman2015}, where the relative error for this critical exponent was $\sim 5\%$ over this same range.  Exact diagonalization (also in Ref.~\cite{Bridgeman2015}) gave a comparable error to MERA.

The estimate $\nupred$ became an increasingly inaccurate predictor of $\nucrit$ towards the ends of the critical line.  We note that, in a previous study of the critical behavior of the quantum AT model using MERA~\cite{Bridgeman2015} (which is also a form of real-space RG), the predicted critical index associated with this exponent also deviated upwards from the exact value towards the endpoint $\lambda=1$ of this critical line, in a way that is very similar to the results shown here.  Exact diagonalization calculations produced a similar behavior.   

As noted previously, this simple approach to real-space RG can accurately reproduce some of the critical exponents of a model, but not necessarily all of them~\cite{Kubica}.  For the Ising model, the scheme of Ref.~\cite{Miyazaki2011} accurately predicts the correlation length exponent, but not the magnetic exponent.  In the quantum AT model, there is a rich critical behavior associated with the scaling dimensions of the orbifold boson CFT, included twisted scaling dimensions that do not appear in the standard $c=1$ boson CFT.  Although such critical indices are not observed here (the lowest such index is $1/8$, independent of $\lambda$), the success in MERA for identifying these additional critical indices~\cite{Bridgeman2015} gives hope for generalizations of simple real-space RG structures such as the one presented here to predict further aspects of the critical behavior.

The SRS RG method can yield a simple, analytical RG map based only on the symmetry properties of the model and yet can also qualitatively reproduce aspects of the critical behavior.  As such, it may be useful as a simple approximation method even if its accuracy is not as competitive as existing numerical tools.  In particular, it may be useful is assessing the role of different symmetries that may be applied to a class of models~\cite{Singh}.  We note that although the quantum AT model studied here is characterised by an Abelian $Z_2 \times Z_2$ symmetry, the SRS RG applies quite directly to continuous symmetries such as SU(2).  While the intertwining requirement of our RG map given in Eq.~(\ref{eqn:SymmetryCondition}) can give a much broader range of solutions for such continuous symmetries, we note that for characterising zero-temperature phases for spin chains with continuous symmetries such as SU($k$) and SO($2k+1$), it is sufficient to characterise the representations of an Abelian subgroup (in these cases, the subgroups $(Z_k \times Z_k)$ and $Z_2^{2k}$ respectively)~\cite{Else}.   

\begin{acknowledgements}
  We thank Jacob Bridgeman for discussions.  We acknowledge support from the ARC via project number DP130103715, the Centre of Excellence in Engineered Quantum Systems (EQuS), project number CE110001013, and the US Army Research Office grant numbers W911NF-14-1-0098 and W911NF-14-1-0103. STF also acknowledges support from an ARC Future Fellowship FT130101744.
\end{acknowledgements}

\end{document}